\documentclass[amsmath,amssymb, aps, prl, twocolumn]{revtex4-2}

\usepackage{float}
\usepackage{graphicx}
\usepackage{dcolumn}
\usepackage{bm}
\usepackage{hyperref}
\usepackage{color}
\usepackage{fancyhdr}
\usepackage[normalem]{ulem}

\definecolor{greenishred}{rgb}{0.6, 0.4, 0}
\newcommand{\comment}[1]{{\color{black}#1}}

\begin{document}

\title{Dissipative Distillation of Supercritical Quantum Gases}

\author{Jorge Mellado Mu\~noz}
\author{Xi Wang}
\author{Thomas Hewitt}
\author{Anna U. Kowalczyk}
\author{Rahul Sawant}
\author{Giovanni Barontini}
 \email{g.barontini@bham.ac.uk}
\affiliation{School Of Physics and Astronomy, University of Birmingham, Edgbaston, Birmingham, B15 2TT, UK}

\date{\today}

\begin{abstract}
We experimentally realize a method to produce non-equilibrium Bose Einstein condensates with {condensed fraction} exceeding those of equilibrium samples with the same parameters. To do this, we immerse an ultracold Bose gas of $^{87}$Rb in a cloud of $^{39}$K with substantially higher temperatures, providing a controlled source of dissipation. By combining the action of the dissipative environment with evaporative cooling, we are able to progressively distil the non-equilibrium Bose-Einstein condensate from the thermal cloud. We show that by increasing the strength of the dissipation it is even possible to produce condensates above the critical temperature. We finally demonstrate that our out-of-equilibrium samples are long-lived and do not reach equilibrium in a time that is accessible for our experiment. Due to its high degree of control, our distillation process is a promising tool for the engineering of open quantum systems.
\end{abstract}

\maketitle

Although ubiquitous in physics, dissipation is usually considered a detrimental mechanism, as it can hinder or interfere with the behaviour of the system under investigation. Notable examples are the friction that limits the performance of classical engines or the decoherence that destroys purely quantum effects. Recently it has been however realized that if properly tamed, dissipation can be used to generate new states of matter~\cite{nicolis1986dissipative, Dogra_2019_3body, buca2019dissipation,dogra_dissipation-induced_2019}, manipulate qubits~\cite{barreiro2011open}, engineer decoherence-free subspaces~\cite{Cirac1995_quantum, Lidar_1998_decoherencefree, Beige_2000_quantum}, generate entangled quantum states~\cite{barontini2015deterministic} and distil quantum features~\cite{Vollbrecht_2011_entanglement}. In particular, when used to drive a system out of equilibrium, dissipation can help in reaching regions of the parameter space that are not accessible to systems in equilibrium~\cite{gaunt_superheated_2013, szymanska2013non}. In the last decades, a large effort has been put in understanding how non-equilibrium many-body systems are created and how they evolve~\cite{Polkovnikov_2011_Nonequilibrium,Langen_2015}. In particular, the tools developed for ultracold atoms have made it possible to experimentally study the dynamics of a wide range of non-equilibrium systems including low dimensional Bose gases~\cite{kinoshita_quantum_2006, hofferberth_non-equilibrium_2007, trotzky_probing_2012}, quenched quantum gases~\cite{sadler_spontaneous_2006, gaunt_superheated_2013, Smith_2012_condensation} and prethermalized states~\cite{gring_relaxation_2012, Langen_2015}.

\begin{figure}[t]
  \centering
  \includegraphics[width=0.49\textwidth]{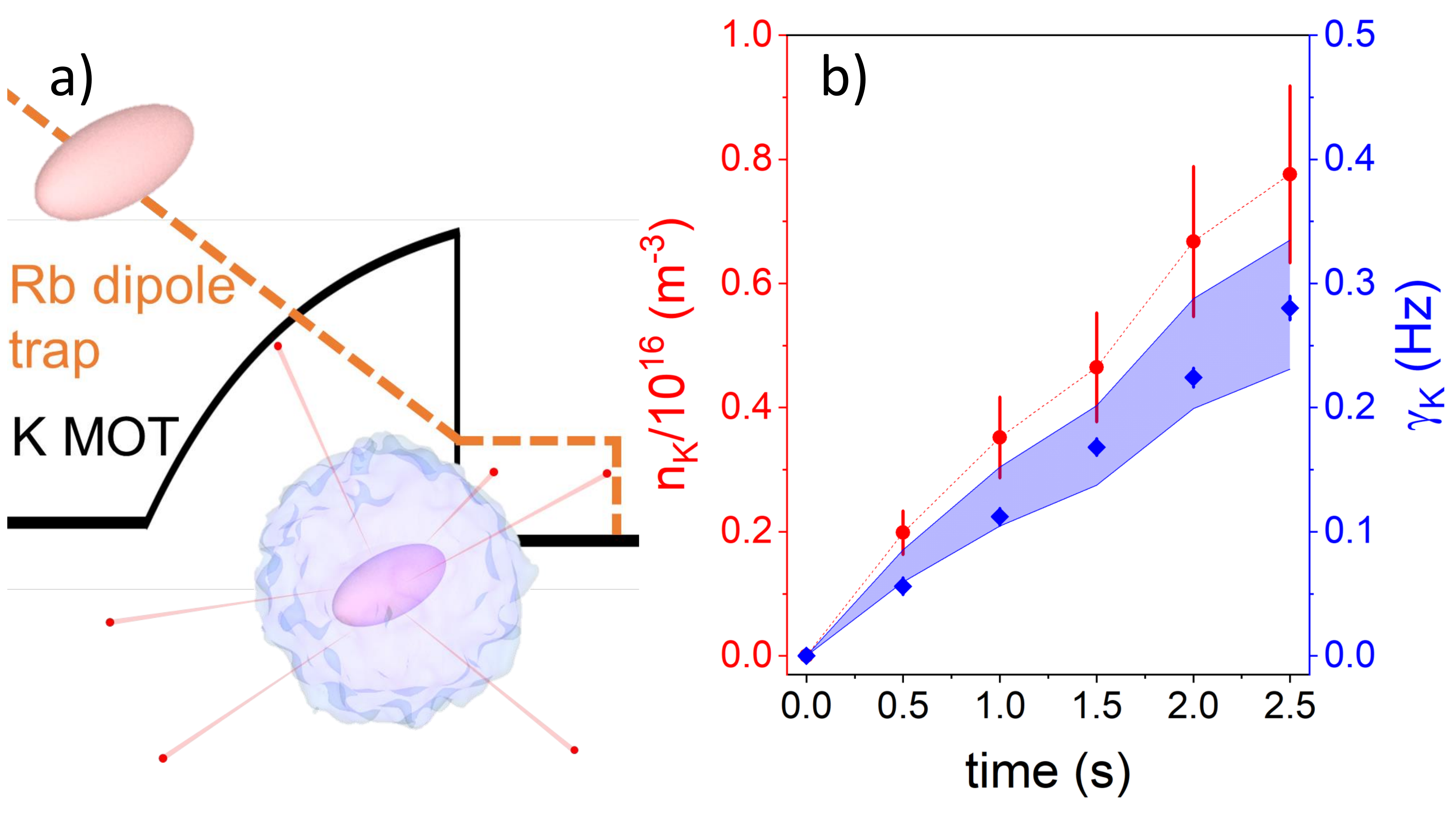}
  \caption{a) Schematic representation of the distillation protocol. During the last stages of the $^{87}$Rb evaporative cooling, we switch on a MOT of $^{39}$K. This results in the creation of a controlled dissipative environment for the Rb atoms. b) The red dots are the measured densities of the $^{39}$K atoms in the MOT as a function of time. The blue diamonds are the corresponding measured dissipation rates while the blue shaded area is the dissipation rate calculated with the model explained in the text. Error bars are the standard errors of the mean.}
  \label{experimental_sequence}
\end{figure}

In this Letter, we study the creation of supercritical non-equilibrium Bose Einstein condensates (BECs) by combining the action of a dissipative environment with evaporative cooling, a process that we refer to as \emph{distillation}. To this end, we immerse an ultracold cloud of $^{87}$Rb at temperatures below 500 nK, within a Magneto Optical Trap (MOT) of $^{39}$K atoms at a temperature of $\simeq$ 1 mK. This causes a loss of Rb atoms with a rate that can be controlled (Fig. 1). We find that the distillation produces long-lived out-of-equilibrium states where the condensed fraction is significantly above the equilibrium value, and even allows us to realize BECs at temperatures higher than the critical temperature. In addition, we show that the distillation prepares the system into quasi-stationary non-equilibrium states that do not reach equilibrium in a time that is accessible for our experiment, therefore exhibiting the features of prethermalized states.

{For an interacting Bose gas in equilibrium in a three-dimensional harmonic trap, the condensed fraction as a function of temperature obeys \cite{BEC_review_Dalfovo_1999, Saturated}:
\begin{equation}
\label{eqn_eql}
F=\frac{N_0}{N}=\begin{cases}
1-\tau^3- \eta \tau^2 (1-\tau^3)^{2/5}  & \text{for} \ T< T_c \\  
0 & \text{for} \ T>T_c \end{cases}
\end{equation}
where $T$ is the temperature of atomic cloud, $T_c=0.94\hbar\omega N^{1/3}/k_B$ is the critical temperature, $\tau = T/T_c$, $N_0$ the number of atoms in the BEC, $N$ the total number of atoms, $\omega$ the geometric average of the trapping frequencies and $\eta = 2.15(aN^{1/6}/a_\text{ho})^{2/5}$, with $a$ the s-wave scattering length, $a_{ho} = (\hbar/m \omega)^{1/2}$ the harmonic trap length and $m$ the mass of the atoms.
}   
In general, in dilute gas experiments neither $\omega$ nor $N$ are constant as the evaporative cooling removes the more energetic atoms from the trap.  In optical dipole traps, this also implies a reduction of the trapping frequencies~\cite{benedicto2017endoscopic}. This in turn leads to lower values of $T_c$ as the evaporation proceeds. Both the reduction of $N$ and $\omega$ are however very mild at the end of an optimized evaporation and Eq.~(\ref{eqn_eql}) is usually valid with very good approximation for the vast majority of experiments \footnote{Further corrections taking into account inter-particle interactions and finite-size effects are discussed in the supplementary materials \cite{SuppMat}}. In general, from Eq.~(\ref{eqn_eql}) it follows that dissipating atoms from the system should result in a reduction of $F$, at least for a cloud in thermal equilibrium. However, this might not be the case if the gas is brought out-of-equilibrium. Under certain conditions, it might indeed happen that a quench in some of the system's parameters leads to long-lived metastable states where $F$ exceeds the value predicted by Eq.~(\ref{eqn_eql}), like e.g. in \cite{gaunt_superheated_2013} where a superheated BEC was realized.

In our experiment, we collect and pre-cool the atoms in a two-species 2D MOT of Rb and K. Using a bi-chromatic beam of light, we then push the atoms from the 2D MOT chamber into the science chamber, where we load the overlapping 3D MOTs of Rb and K. Typically, we trap and cool $\approx 10^9$ Rb atoms at a temperature of 300~$\mu$K and $\approx 10^7$~K atoms at a temperature of 1~mK. For the experiments described in this Letter, we start by loading only the Rb 3D MOT. We subsequently load the Rb atoms directly from the MOT into an optical dipole trap formed by crossing, at an angle of $\simeq$~40 degrees, two beams of wavelength 1070~nm and 1550~nm, with waist sizes of 35~$\mu$m and 45~$\mu$m respectively. Once the atoms are loaded in the dipole trap, we switch off the MOT magnetic field gradient and beams, and evaporatively cool the atoms down to the degenerate regime in 10 s. In the last 6 s of the evaporation we switch on again the MOT magnetic field gradient and obtain a BEC with $3 \times 10^4$ atoms in the $|F=1, m_F=-1\rangle$ state. The final trapping frequencies are \comment{$\simeq2\pi\times(70,120,120)$ Hz}. Unless otherwise stated, at the end of the sequence we hold the atoms for 20 ms in the dipole trap before releasing them and taking absorption images in time-of-flight.

As shown in Fig.~\ref{experimental_sequence}, to immerse the ultracold Rb gas in a dissipative environment, we switch on the K MOT during the last stage of the evaporation, when the Rb temperature is below 1 $\mu$K, for a variable amount of time. To this end it is sufficient to switch on the K push and MOT beams, as the quadrupole magnetic field is already on. In Fig.1b we report the growth of K atom density $n_K$ as a function of the loading time (red circles). In the same figure, we report the corresponding dissipation rate $\gamma_K$ as measured in our experiment (blue diamonds). As the temperature of the K atoms is $\approx 1$~mK, more than three orders of magnitude higher than the temperature of the Rb gas and the dipole trap depth, most of the collisions between K and Rb lead to the loss of Rb atoms from the dipole trap. Indeed the measured $\gamma_K$ coincides with the value obtained with $\gamma_K=n_K \sigma v_K$ (shaded area), where $v_K$ is the average speed of the K atoms and $\sigma$ is calculated using the model of \cite{cornell1999experiments} for collisions between ultracold atoms and background classical atoms \cite{SuppMat}. For comparison, the Rb elastic scattering rate $\gamma_\textrm{el}$, that is responsible for the thermalization of the Rb cloud, ranges between $\simeq$~10-65~Hz for the experiments here reported.  

\begin{figure}[t]
  \centering
  \includegraphics[width=0.49\textwidth]{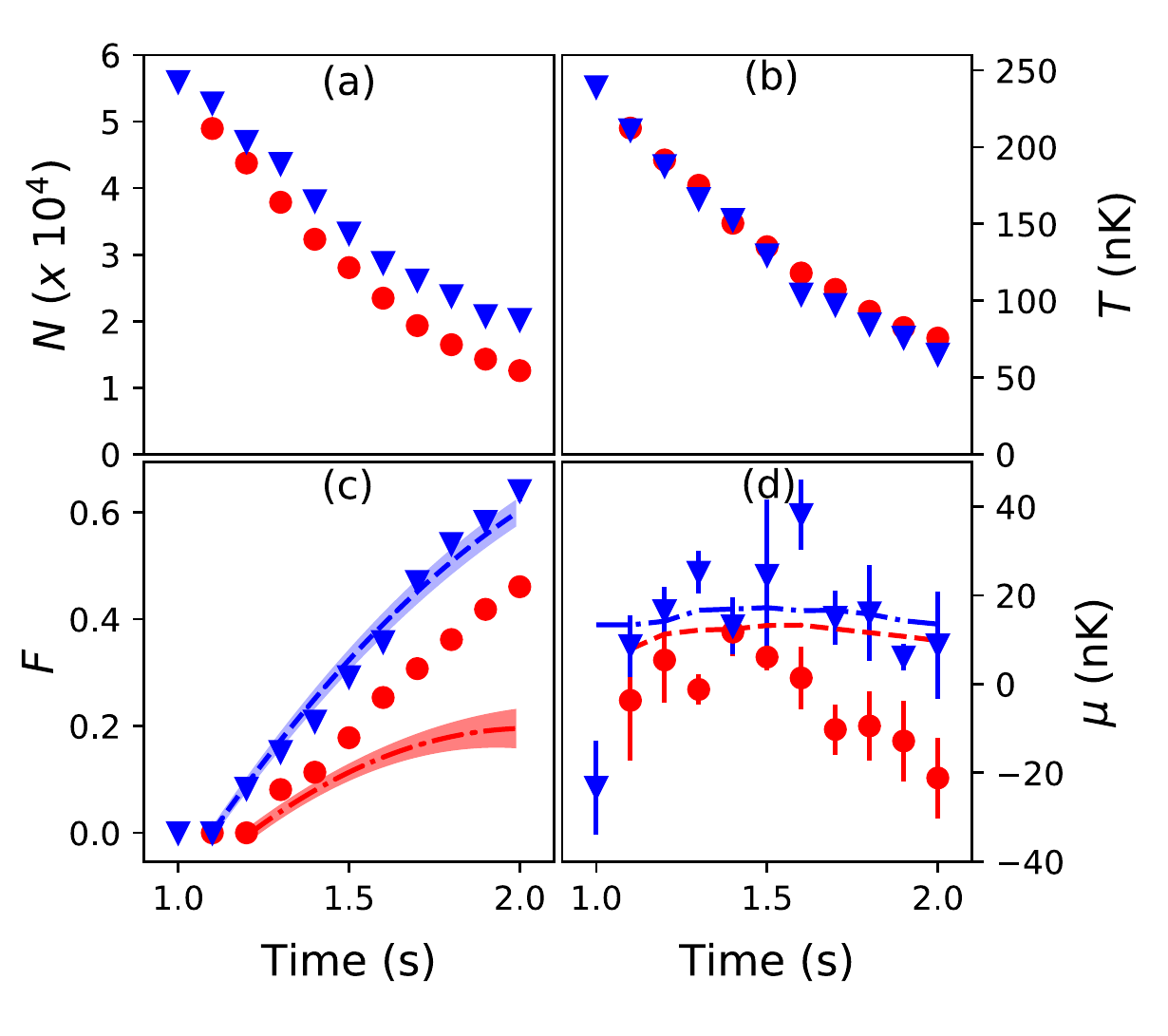}
  \caption{Measured parameters of the $^{87}$Rb sample as a function of time. Red circles are with dissipation (distillation) and blue triangles are without dissipation. a) The total number of atoms.
  b) The temperature c) The condensate fraction $F$. The red dotted dashed line and the blue dashed line show the expected fraction from Eq.~(\ref{eqn_eql}) for with and without dissipation respectively \cite{SuppMat}. d) The chemical potential of the thermal part in units of temperature. The dashed and dash-dotted lines are the chemical potential of the BEC with and without dissipation respectively. Error bars and the shaded regions are the standard errors of the mean.}
  \label{all_graphs}
\end{figure}

In Fig.~\ref{all_graphs} we report the typical temporal evolution of the parameters of the Rb gas across the BEC transition with and without the dissipation. For the reported data, the K MOT is switched on 2 s before the end of the evaporation, where we set $t=0$. For a direct comparison, the reported data without dissipation are chosen to approximately match the conditions with dissipation at $t=1$ s, right before the onset of the BEC. As expected, when the dissipation is present we observe that the number of atoms is decreasing at a faster rate than the optimized evaporation (Fig.~\ref{all_graphs}a). Crucially, the evaporation selectively removes only the more energetic atoms from the cloud, while the dissipation coming from the K MOT is uniform and acts equally on all the velocity classes. This is reflected also in the behaviour of the temperature (Fig.~\ref{all_graphs}b), {which does not change substantially when the dissipation is present} \cite{footnote}. It also confirms that the action of the K MOT is purely dissipative (no heating) and that the dissipation does not affect the evaporative cooling and the ability of the Rb cloud to rapidly thermalize. 

The corresponding measured condensed fraction $F$ as a function of time is shown in Fig.~\ref{all_graphs}c. We observe that in the presence of the dissipation this is significantly higher than what is predicted by Eq.~(\ref{eqn_eql}) (dashed curve). Notably, as the distillation proceeds, the discrepancy between the measured $F$ and that predicted by Eq.~(\ref{eqn_eql}) increases, producing a BEC substantially more pure than what can be obtained with the same atom number and temperature but without dissipation \cite{SuppMat}.
Fig.~\ref{all_graphs}d finally shows how the chemical potential of the non-condensed part of the cloud $\mu$ changes differently for with and without dissipation~\cite{SuppMat}. 
As expected, for both cases $\mu$ initially approaches the chemical potential of the BEC (lines). However, with distillation the behaviour of $\mu$ is non-monotonic and above $\simeq1.5$ s reduces even when $F$ increases, creating a system which is not in phase equilibrium~\cite{gaunt_superheated_2013}.

\begin{figure}
  \centering
  \includegraphics[width=0.49\textwidth]{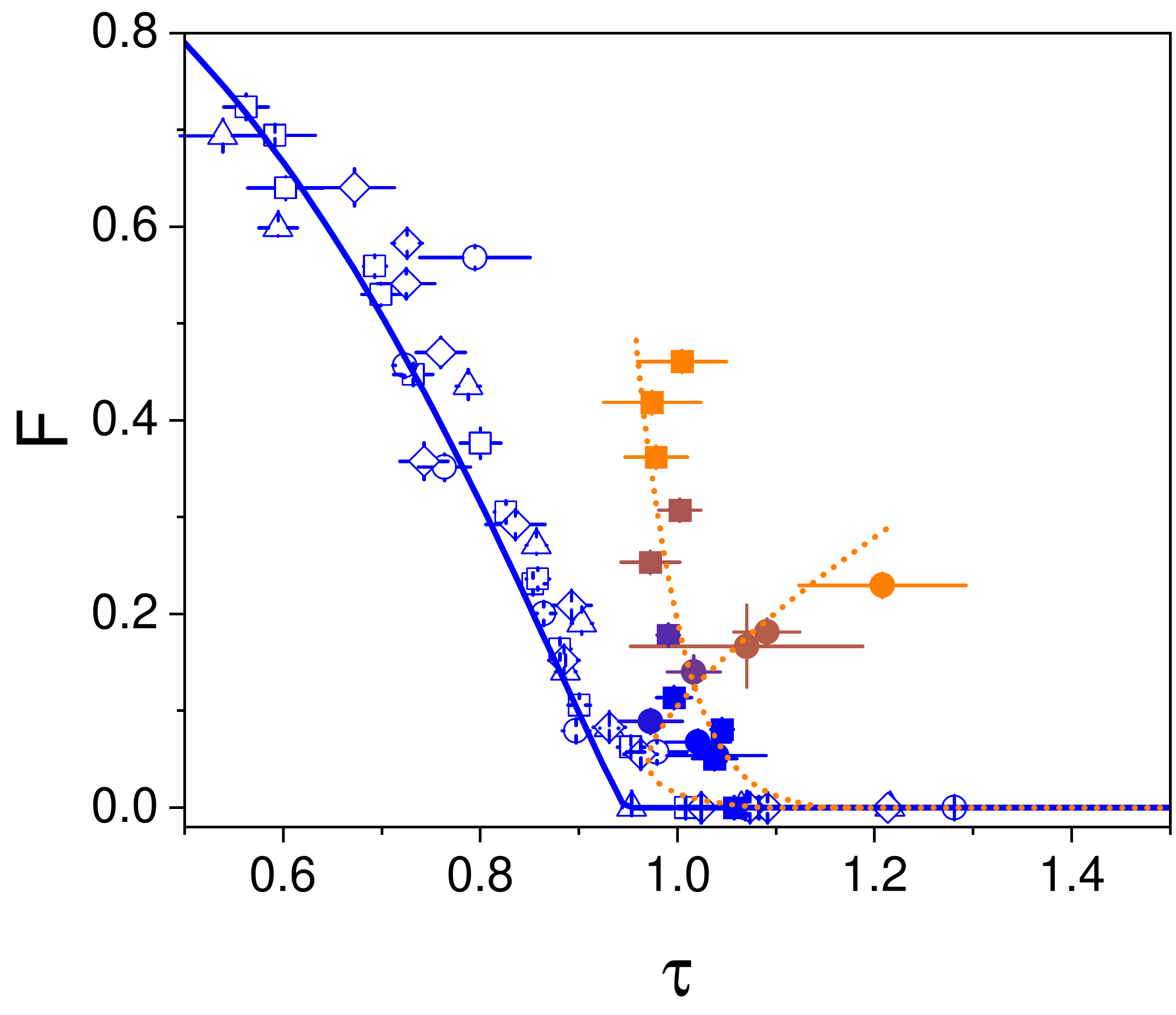}
  \caption{Phase diagram showing the BEC fraction as a function of $\tau$. Open symbols are for the case of no dissipation while filled symbols are the trajectories during our distillation protocol. For open diamonds and circles the wait time after the end of the evaporation ramp is 4 s, while for others it is 20 ms. For the filled circles the dissipation rate is 4 times the dissipation rate of the filled squares. {For the data in absence of dissipation, we vary the number of atoms from 70$\times 10^3$ to 25$\times10^3$ at $t=1$s} in order to explore as much parameter space as possible. {The solid blue line corresponds to Eq.~\ref{eqn_eql}}.The dotted lines are the results of the rate equation model described in the text \cite{SuppMat}. Error bars are the standard errors of the mean.
  }
  \label{phase_diagram}
\end{figure}

In Fig.~\ref{phase_diagram} we report the data as trajectories in the $F-\tau$ plane. The open symbols are the results that we obtain without dissipation, varying the initial conditions or the hold time at the end of the evaporation. The solid blue curve corresponds to Eq.~(\ref{eqn_eql}). This demonstrate that our optimized evaporation produces samples in equilibrium over a broad range of initial conditions, and that we don't need long hold times at the end of the sequence to reach equilibrium. The filled squares in Fig.~\ref{phase_diagram} correspond instead to the data reported in Fig.~\ref{all_graphs}. When the dissipation is switched on, the trajectory substantially differs from Eq.~(\ref{eqn_eql}), and notably we are able to progressively distil purer samples. Our distillation allows us to explore regions of the phase diagram that are not accessible for gases in equilibrium and that feature a higher purity. We refer to those samples as \emph{supercritical BECs}. 

During the distillation, the dissipation shifts $T_c$ to lower values, counter-acting the action of the evaporation that reduces $T$, so that $\tau$ remains approximately constant. However, at the same time $F$ increases, meaning that while the reduction of $T$ pumps atoms in the BEC, the reduction of $T_c$ is not able to de-pump them back into the thermal component at the same rate. As it can be observed in Fig.~\ref{phase_diagram}, the result is a steeper purification with distillation, and a BEC with $F\simeq 0.5$ can be produced already for $\tau\simeq 1$. This effect is even more apparent if we increase the rate of dissipation by a factor of 4 (filled circles). To do so, we increase the power of the push beam, so that the loading rate of the K MOT is quadrupled. In this case, the distillation is so effective that the trajectory \emph{inverts} and we are able to increase $F$ even if we increase $\tau$.

\begin{figure}
  \centering
  \includegraphics[width=0.4\textwidth]{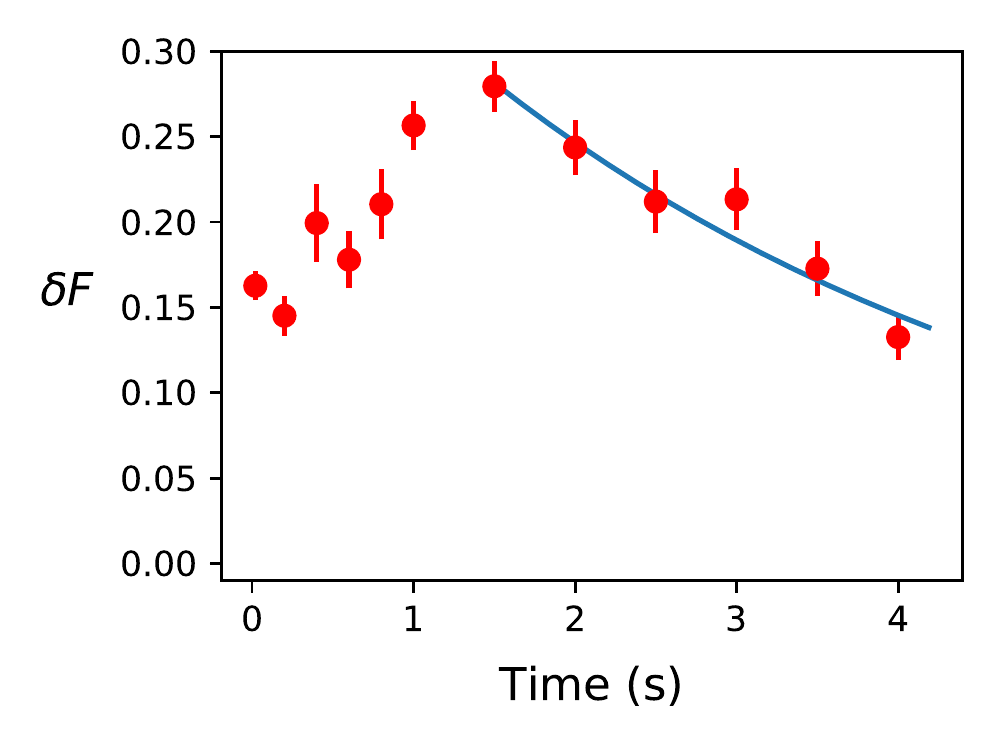}
  \caption{Difference between the measured $F$ and that expected using Eq.~(\ref{eqn_eql}) after the dissipation from the K MOT is switched off and the sample is held with constant trap depth. Error bars are the standard errors of the mean. The line shows an exponential fit to the data after 1.5 seconds, when the supercritical gas relaxes towards equilibrium. The time constant is {$3.9\pm0.3$ s}.}
  \label{relaxation_fig}
\end{figure}

In Fig.~\ref{relaxation_fig} we address the issue of the lifetime of our supercritical states. To measure the lifetime, we switch off the dissipation right after the state has been created following a trajectory similar to the one of Fig. 2. Then we keep the cloud in the dipole trap with a constant trap depth for a variable amount of time. In Fig.~\ref{relaxation_fig} we report the difference $\delta F$ between the measured $F$ and Eq.~(\ref{eqn_eql}) as a function of time after the dissipation has been switched off \cite{SuppMat}. For the first 1.5 s, the system is driven even further out of equilibrium by plain evaporation and then it slowly relaxes toward lower values of $\delta F$. However, for as long as we can measure, $\delta F$ never goes below the initial value. 
With respect to the typical timescales of the experiment, {which range from $1/\omega\simeq 0.1$~ms to $1/\gamma_{el}\simeq100$ ms}, the relaxation dynamics can therefore be considered quasi-static, meaning that our supercritical samples possess similar properties as a prethermalized state.

The dynamics of the formation of the BEC during evaporative cooling is a complex many-body problem and a microscopic theory able to quantitatively describe it still doesn't exist. Some models have tried to reproduce the experimental observations, but only for peculiar settings (constant temperature and infinite atom reservoir) and with partial success \cite{Jaksch_1997,Gardiner_1998}. The addition of the dissipation makes the microscopic description of our dissipative distillation an even more challenging task. We have however developed a phenomenological rate equation model starting from those in  \cite{Jaksch_1997,Gardiner2,Gardiner_1998,gaunt_superheated_2013}. This allows us to describe our experimental data and derive important information that can be used to develop a rigorous microscopic theory. Our model describes our system as a two-mode system, with one mode being the BEC and the other the thermal component \cite{SuppMat}:
\begin{align}
\dot{N}_0&=\bar{W}\left[\left(1-\frac{t}{t_f}\right)N_0+1\right]-\bar{K}(\tilde{N}_{th}+1)-\gamma_K(t)N_0 \nonumber \\
\dot{N}_{th}&=-\bar{W}\left[\left(1-\frac{t}{t_f}\right)N_0+1\right]+\bar{K}(\tilde{N}_{th}+1)+\nonumber\\&-[\gamma_K(t)+\gamma]N_{th}. 
\end{align}
$\bar{W}$ and $\bar{K}$ are respectively the growth rate of the condensate and of the thermal component and are derived from the data without dissipation. The loss rate $\gamma$ accounts for the evaporative cooling while $t_f$ for the saturation of the BEC, also these parameters are extracted from the data without dissipation. $\tilde{N}_{th}$ is the effective number of atoms in the thermal mode and is the only free parameter of our model \cite{SuppMat}. The results are reported as dotted lines in Fig. 3 where it can be appreciated that our model is able to reproduce fairly well the trajectories of our dissipative distillation.

The crucial element of our dissipative distillation is the fact that the rates $\bar{W}$, promoted by the reduction in temperature coming from the evaporative cooling, and $\bar{K}$, promoted by a reduction of the chemical potential coming from the dissipation, do not coincide for a Bose gas out of equilibrium. {By considering two-body collisions as the only mechanism responsible for the growth of the condensate, and using quantum kinetic theory, it is indeed possible to demonstrate that $\bar{W}\simeq \exp(\Delta/k_BT)\bar{K}$, with $\Delta$ the energy difference between the two components \cite{Jaksch_1997,Gardiner_1998, SuppMat}. The energy gap can be roughly estimated from the energy spectrum obtained with a first-order treatment of a uniform Bose gas with contact interactions. For $\tau\leq1$, this reduces to \cite{Huang_quantum_1957,Huang_imperfect_1957}:
\begin{equation}
    E=\sum_p \frac{p^2}{2m}+\frac{4\pi\hbar^2 aN^2}{mV}\left(1-\frac{1}{2}F^2\right)
\label{eq2}    
\end{equation}
where $p$ is the momentum of the atom, $m$ its mass, $a$ the s-wave scattering length and $V$ the trapping volume. The last term is of quantum mechanical origin and accounts for bosonic stimulation. From eq. (\ref{eq2}) it follows that once an atom is in the condensed phase, it needs an amount of energy $\Delta=2\pi\hbar^2 aN_0/mV$ to leave the BEC, yielding an unbalancing between $\bar{W}$ and $\bar{K}$. More detailed calculations including higher order perturbation theory \cite{Huang_quantum_1957,Huang_imperfect_1957} and the effect of the mean field potential of the BEC \cite{Gardiner_1998,Gardiner2} show that the spectrum exhibits a strong modification of the density of states right above the condensed state, therefore Eq. (\ref{eq2}) is valid only for low values of $F$. Regardless, for $\tau\simeq1$, in our experimental conditions $\Delta$ is already of the same order of magnitude as $T$.}

In conclusion, we have implemented an open many-body quantum system by immersing an ultracold gas in a controlled dissipative environment, embodied by a cold gas of atoms of a different species. We have shown that by combining the dissipation with evaporative cooling it is possible to realize states of matter that are not accessible for equilibrium or closed systems. In particular we were able to create and grow supercritical BECs, even at temperatures higher than the critical temperature. The states created exhibit a quasi-static behaviour typical of prethermalized states and can be practically used to perform experiments with high condensed fractions at high temperatures. On the one hand, our results have the potential to trigger the interest of the theory community to develop a microscopic description of out-of-equilibrium quantum gases. On the other, the ability to control the dissipation and the temperature of the sample can provide a new tool for distilling environmentally resilient states and engineering quantum phases in open quantum system.

\paragraph*{Acknowledgements}
The authors are supported by the Leverhulme Trust Research Project Grant UltraQuTe (grant number RGP-
2018-266). We are grateful to J. Goldwin and V. Boyer and V. Guarrera for reading the manuscript and the useful comments. We acknowledge fruitful discussions with the members of the Cold Atoms Group at the University of Birmingham.

\bibliography{main_bibl}

\begin{thebibliography}{35}
\bibitem [{foo()}]{footnote}%
  \BibitemOpen
  \href@noop {} {}\bibinfo {note} {The temperature of the data with dissipation
  is systematically higher than in absence of dissipation, this is because the
  evaporative cooling is less efficient as the number of atoms is
  reduced.}\BibitemShut {Stop}%
\end{thebibliography}

\end{document}


\title{Supplementary material for: \\ Dissipative Distillation of Supercritical Quantum Gases}

\author{Jorge Mellado Mu\~noz}
\author{Xi Wang}
\author{Thomas Hewitt}
\author{Anna U. Kowalczyk}
\author{Rahul Sawant}
\author{Giovanni Barontini}
 \email{g.barontini@bham.ac.uk}
\affiliation{Midlands Ultracold Atom Research Centre, School Of Physics and Astronomy, University of Birmingham, Edgbaston, Birmingham, B15 2TT, UK}

\maketitle
\section{Calibration of the trapping frequencies}
During the evaporation, the trap depth is reduced slowly. This changes the trapping frequencies (we indicate with $\omega$ the geometric mean of the trapping frequencies) of the dipole trap and hence the critical temperature according to $k_B T_c = 0.94\hbar\omega N^{1/3}$. 

\begin{figure}
    \centering
    \includegraphics[width=0.48\textwidth]{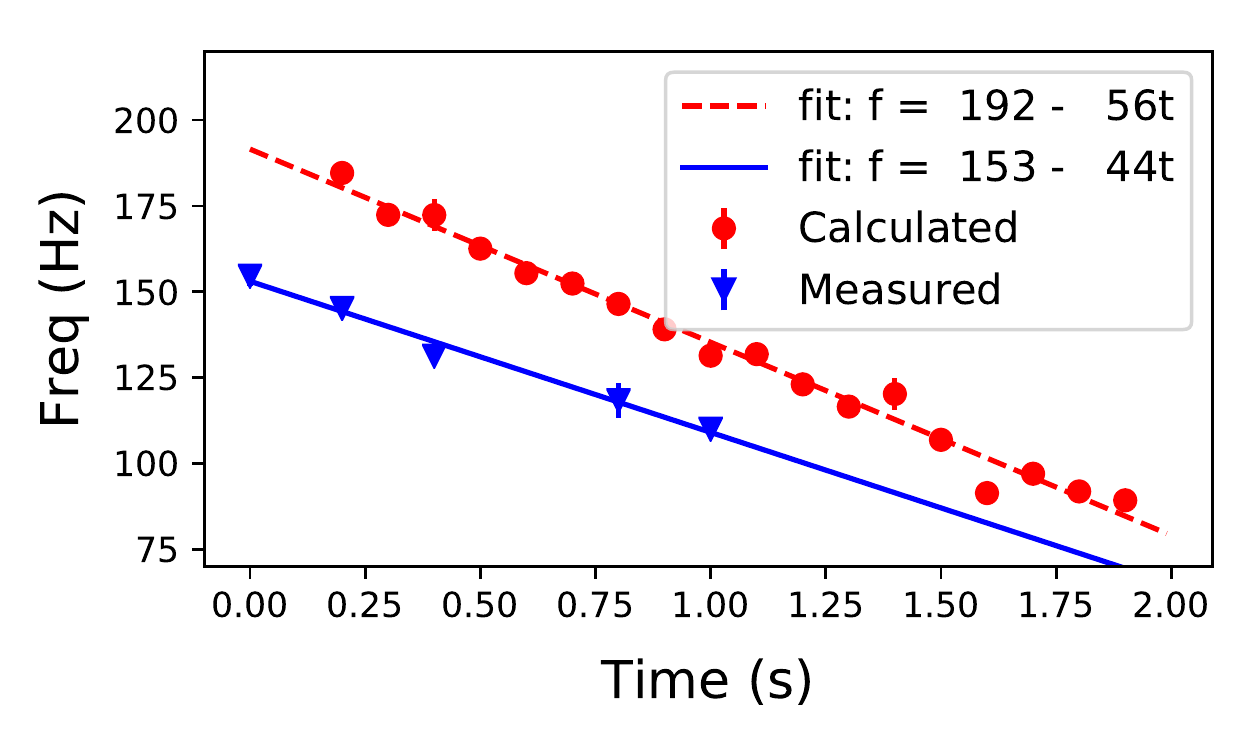}
    \caption{ The red dots are the mean trapping frequency $\omega/2 \pi$ calculated using the method described in the text. The blue triangles are the mean trapping frequencies measured using parametric heating and centre of mass oscillations. The lines are linear fits to the data, the coefficients are reported in the legend. 
    }
    \label{fig:freq_fit}
\end{figure}

In equilibrium, the BEC fraction is {$F = 1 - \tau^3- \eta \tau^2 (1-\tau^3)^{2/5}$. We use this fact to estimate the trapping frequency at each time during the evaporation by equating experimentally observed BEC fraction to this formula and numerically solving for $\omega$.}
The mean trapping frequency calculated using the data of no dissipation cases from Fig.~3 in the main text is shown in Fig.~\ref{fig:freq_fit}. We verified that all the data without dissipation correspond to equilibrium configurations by waiting a variable amount of time after the evaporation is finished. Data with the same evaporation ramp but different waiting times share the same value of $\omega$. As shown in Fig. 3 in the main text all data collapse on the equilibrium curve independent of the waiting time.

After having estimated $\omega$ we perform a linear fit to the frequency vs. evaporation time data reported in Fig. S1. The values from the linear fit are then used to calculate $T_c$ for each non-equilibrium data point in Figs.~2 and 3 in the main text. 

We also measure the trapping frequencies in a few points along the evaporation ramp. The blue triangles in Fig.~\ref{fig:freq_fit} show the geometric mean of the trapping frequencies measured using parametric heating of the thermal Rb cloud. The measured values are consistent within $\sim 15 \%$ of the values used for calibration. This is expected as it is known that the frequency measurements using parametric heating lead to a systematic underestimation due to trap anhormonicities~\cite{MAKHALOV2015327}. The calibration also accounts for effects of systematic errors in the data like, e.g., a systematic over- or under-estimation of the total number of atoms and common mode secondary effects as discussed in the following sections. The linear drop in frequency along the evaporation approximately matches the frequencies change we expect from the lowering of the power of the dipole trap lasers.

\begin{figure}
    \centering
    \includegraphics[width=1\textwidth]{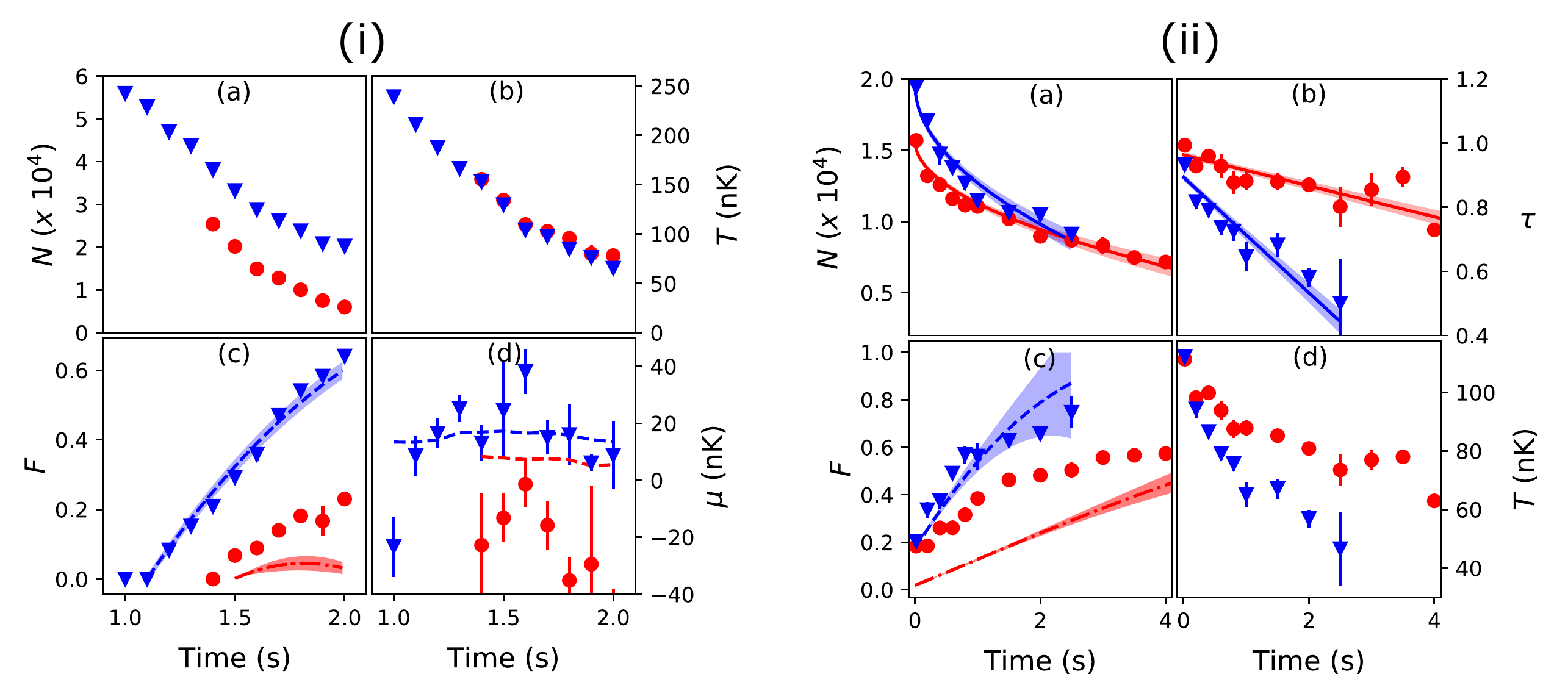}
    \caption{{(i) Same as the figure 2 in the main text, but with the dissipation rate increased by a factor of $\simeq$4, corresponding to filled circles in Fig. 3 of the main text. (ii) Measured parameters as a function of time for the experiment corresponding to Fig.~4 (in the main text). For Red circles dissipation was present before the zero time and for blue triangles there was no dissipation at any point in time. ii.a) The total number of atoms.  ii.b) The ratio of temperature to the critical temperature. ii.c) The condensate fraction $F$. The red dotted dashed line and the blue dashed line show the expected fraction from the equilibrium theory for with and without dissipation respectively. Here $\tau$ is fitted to a linear function of time and $N$ to $\sqrt{t}$ to get smooth curves for the theoretical prediction. ii.d) The temperature. Error bars are the standard errors of the mean and the shaded regions are from standard errors of fit parameters.}}
    \label{fig:all_dat}
\end{figure}

A similar calibration is also done for Fig.~4 in the main text. Here the frequency does not change in time as we keep the power of the dipole trap lasers constant. Even though the lowering of trap depth is stopped, the system still undergoes plain evaporative cooling and the temperature is slowly reduced for $\simeq2$ s. As the plain evaporation is very slow, the system without dissipation (blue triangles) is always in equilibrium and we can use the data in Fig. \ref{fig:all_dat} to extract $\omega$ and calculate the equilibrium curves for the dissipation case (red circles). The $\Delta_F$ data reported in Fig. 4 in the main text correspond to the difference between the red circles and the red line in Fig. \ref{fig:all_dat}c.

The values for $N_0$ and $T$ are obtained by fitting a combination of two-dimensional Thomas-Fermi and Gaussian distributions to the atomic density image obtained from time of flight absorption imaging. The value of $N$ is obtained by integrating the atomic density obtained from the image.

{\section{Effect of finite size and Bose statistics}}
In the main text, we compare our results against Eq. (1). 
There are however a number of known effects that can lead to deviations from  this equation. 
In this section, we quantify these discrepancies for our experimental conditions. The major effects to be accounted for are: finite size effects, the Bose enhancement effect, changes in the transition temperature due to inter-atomic interactions between Rb atoms. The finite size effect on the transition temperature can be calculated as $\delta T_c/T_c=-\bar{\omega}/2\omega\cdot \zeta(2)/\zeta(3)^{2/3}\cdot N^{-1/3}$~\cite{BEC_review_Dalfovo_1999} with $\bar{\omega}$ the arithmetic average of the frequencies. This contribution never exceeds  $1 \%$ for our system, therefore can be safely neglected. We discuss the other corrections here below.

\subsection{Deviation from Gaussian distribution in momentum space}
As specified above, the fraction of BEC and the temperature of the thermal cloud for the data shown in the main text is obtained by fitting a combination of a 2-D Gaussian and a 2-D Thomas-Fermi distribution to the absorption image of the whole atomic sample. The total number of atoms is obtained by integrating the atomic density from the absorption images. 
However, for $\tau\simeq1$ and below, the Bosonic nature of the Rb atoms changes the momentum distribution of the cloud from Gaussian to the so-called Bose-enhanced distribution. For this reason, to be rigorous, data should be fitted using the associated polylogarithm~\cite{ketterle1999making}. 

\begin{figure}
    \centering
    \includegraphics[width=0.45\textwidth]{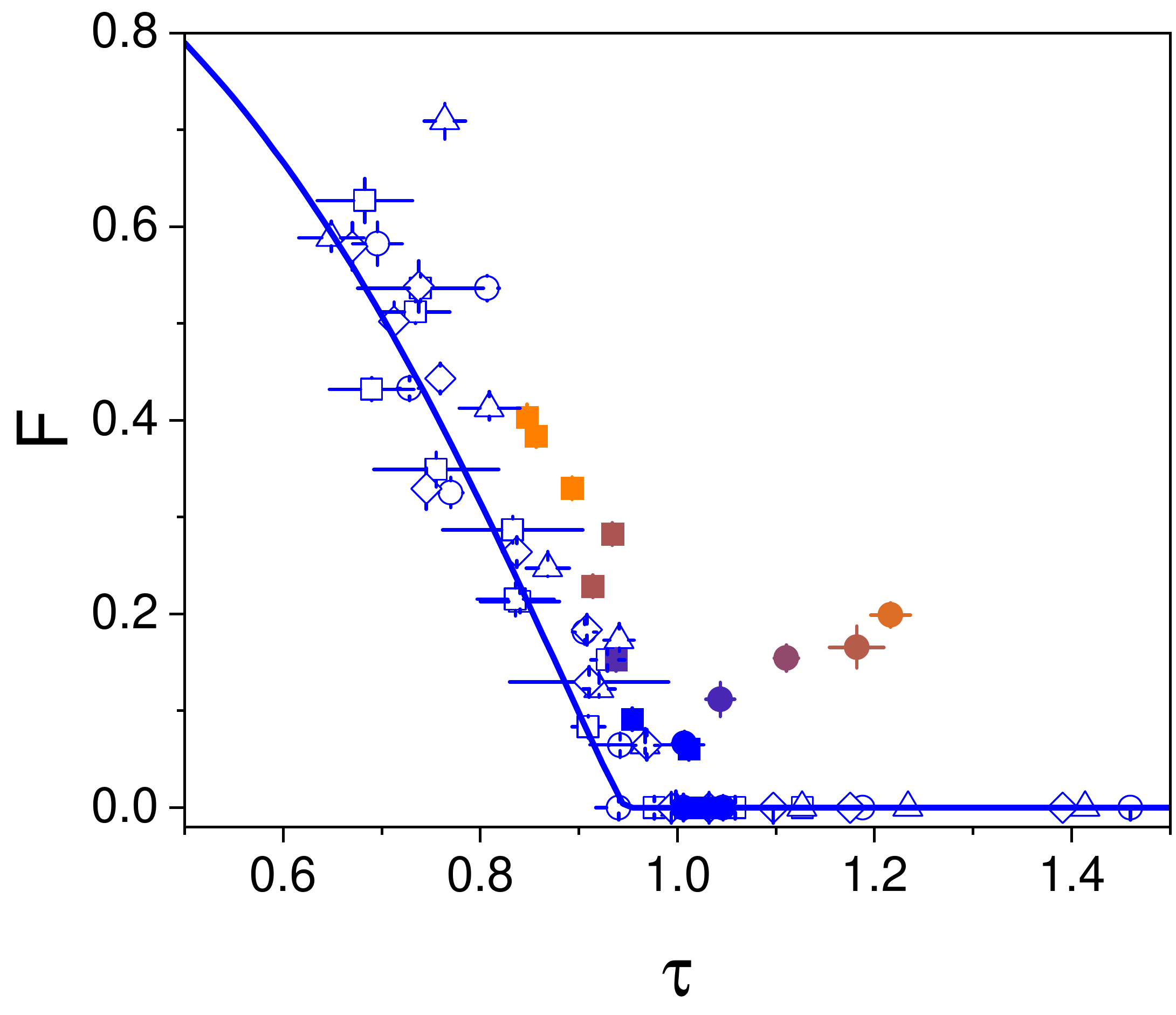}
    \caption{Same as Fig.~3 in the main text but using the Bose-enhanced polylogarithm to fit the data.}
    \label{fig:S3}
\end{figure}

To quantify the effect of the choice of the fitting function, we fit our data also with a combination of a two-dimensional Bose-enhanced Gaussian and a Thomas-Fermi distribution. As shown in the Fig.
\ref{fig:S3}a, there is no qualitative change in the overall behaviour of our data. The main effect is that our estimation of the temperature increases by $\sim 10 \%$ on average. The other major effect is that, for low atom number in the thermal component, the fit function does not always converge due to the complexity of the polylogarithm function. Therefore we cannot use the full set of our data, yielding to an increase of our errorbars and distortion of the trajectories. As discussed above, systematic effects on temperature and atom number are however absorbed in the calibration of the trapping frequencies and do not play a significant role in the physics discussed in the main text. For this reasons, we provide in the main text the data analysed using Gaussian distributions, which is more robust and allows us to use the full set of our data. 

\subsection{Further effects of atomic interactions}
Eq.~(1) from main text is an approximation and is derived by neglecting the interactions within the thermal cloud. This approximation is valid as long as $N_0 \gg a/a_\text{ho}$. For our experiment this corresponds to $N_0 \gg 1000$. This condition holds for a large subset of our experimental data.

The other effect of the interactions is to shift the critical temperature to lower value~\cite{BEC_review_Dalfovo_1999,shift1,shift2}. For our experimental conditions, the shift is $\delta T_c/T_c=-1.3a/a_{ho}N^{1/6}< 3\%$. Like the majority of the effects discussed in this section, the shift is common mode to all the data and is taken care of by our calibration procedure.  

\section{Dissipation rate}
In our system $T_K \simeq 10^4 T_{Rb}$, therefore we use the model of ~\cite{cornell1999experiments} for collisions between ultracold atoms and background atoms to estimate the KRb scattering rate. The model features two regimes, 'quantum' and 'classical', delimited by the crossover energy $E_c$ given by

\begin{equation}
\label{Ecross}
    E_{c} \approx 2\frac{\hbar^{12/5}}{C_{6}^{2/5}}\frac{1}{m_{Rb} m_{K}^{1/5}}E_{col}^{1/5},
\end{equation}

where $m_{Rb}$ and $m_K$ are the masses of the Rubidium and Potassium atoms respectively. $E_{col}$ is the collision energy, which given the difference in temperature is essentially the kinetic energy of the K atoms, and $C_6$ is the van der Waals coefficient, that for KRb is $C_6$=4290(2)$a_0^6 E_h$, where $E_h$ is the Hartree Energy~\cite{simoni2008near}. Using eq.~(\ref{Ecross}) we calculate $E_c\simeq300\mu$K. The differential cross section, $d \sigma_K$, is given by the expression
\begin{equation}
    d \sigma_K(E_t)=\left\{ 
                \begin{array}{lll}
                  \alpha & \text{if} & E_t \leq E_c  \\
                  \alpha \left( \frac{E_c}{E_t} \right)^{7/6} & \text{if} & E_t > E_c   \\
                \end{array}
              \right.
              \label{intr}
\end{equation}
where $E_t$ is the energy transferred to the Rb atom and $\alpha=\sigma_{g}^{class}(E_c)$, with
\begin{equation}
    \sigma_{g}^{class}(E_t) = \frac{\pi}{6} \left(\frac{9C_{6}^{2}}{E_{col}} \right)E_{t}^{-7/6}.
\label{Scatclass}
\end{equation}
By integrating \ref{intr} using our experimental parameters we obtain that the total scattering cross section for the KRb collisions is $\sigma_K \simeq 5.61 \times 10^{-17} m^2$. Using this value of $\sigma_K$ we can calculate the scattering rate $\gamma_K$ as
\begin{equation}
    \gamma_K = n_K \sigma_K v_K, 
\end{equation}
where $n_K$ is the K atom density and $v_K$ is the average speed of the K atoms. 

\begin{figure}
    \centering
    \includegraphics[width=0.3\textwidth]{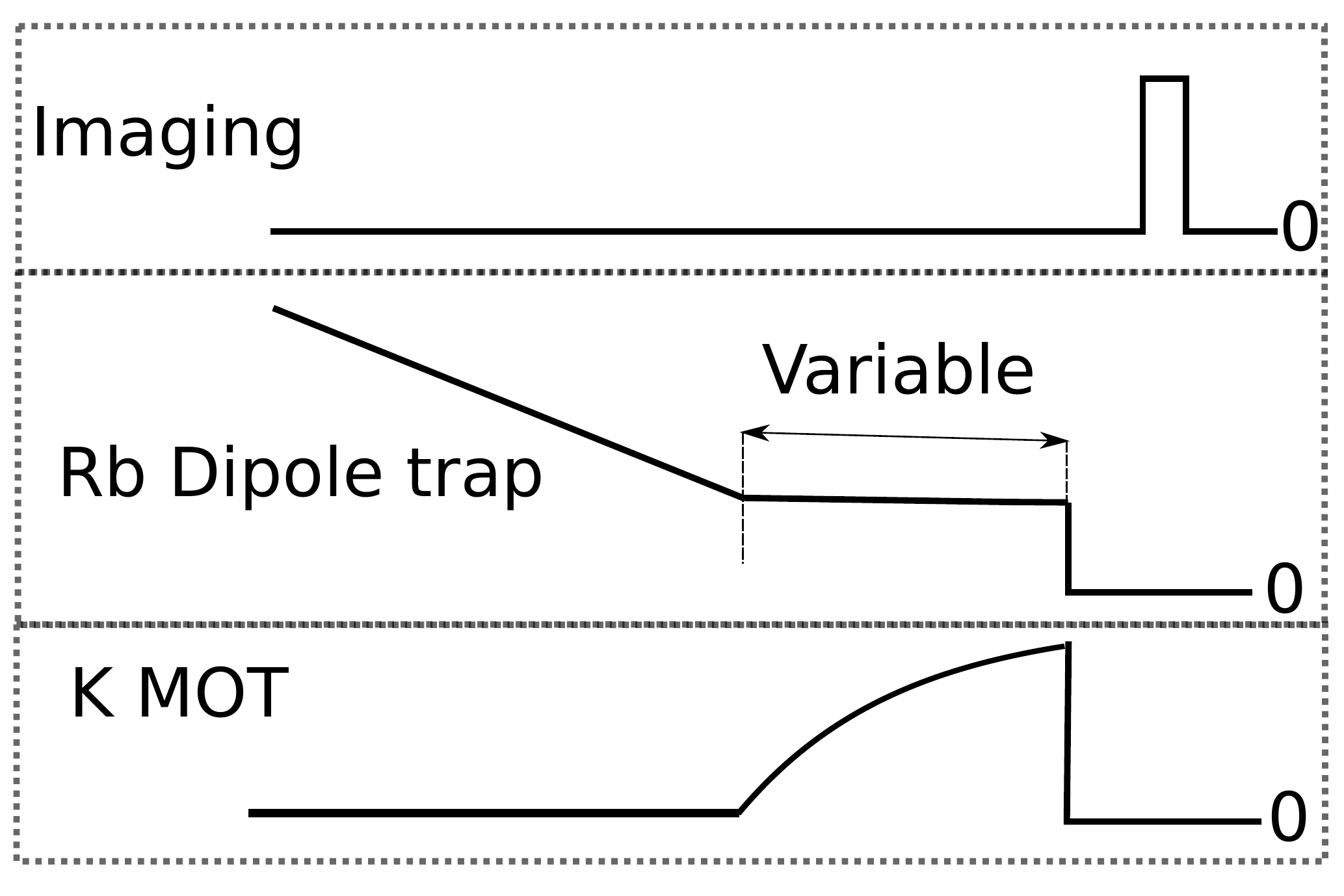}
    \caption{Experimental sequence used to determine the
    dissipation rate of the Rb atoms due to the K atoms.}
    \label{fig:sequence}
\end{figure}

To obtain the experimental value for $\gamma_K$ instead, we use the sequence shown in Fig \ref{fig:sequence}. We switch on the K MOT in the moment we stop the evaporation and keep the dipole trap power at a constant value. Therefore, we can write the following expression to describe the loss of Rb atoms,
\begin{equation}
     \dot{N}_{Rb} = -( \gamma_{Rb} + \gamma_K (t)) N_{Rb},
    \label{rate} 
\end{equation}
where $\gamma_{Rb}$ is the loss rate for the Rb atoms in the dipole trap due to the combination of plain evaporative cooling and three-body losses. In our case $n_K$ grows approximately linearly in time so $\gamma_K (t)=at$ and
\begin{equation}
    N_{Rb} (t) = N_{Rb} (0) \exp \left(-\gamma_{Rb} t - \frac{a}{2} t^2   \right).
\end{equation}

\begin{figure}
    \centering
    \includegraphics[width=0.4\textwidth]{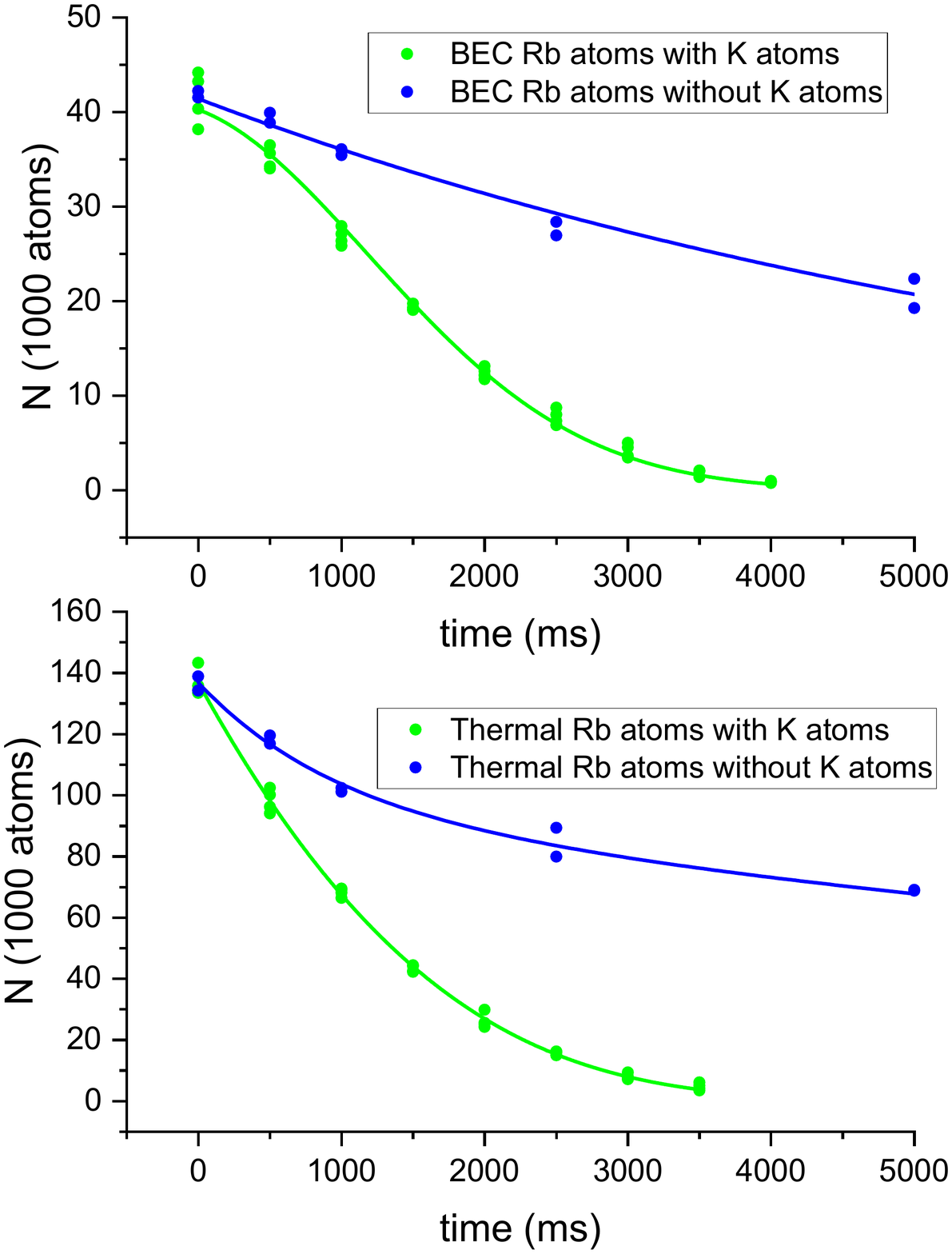}
    \caption{Rb atom number as a function of time for a BEC sample (upper panel) and a thermal gas (lower panel). The blue points and curves are without the dissipation from the K MOT. The green points and curves correspond to the case with the K MOT on.}
    \label{fig:my_label5}
\end{figure}

In Fig.~\ref{fig:my_label5} we show the data and corresponding fits for $N_{Rb}$ as a function of time for a pure BEC and a thermal cloud with no BEC component. In both cases we use the measurements without the K MOT to obtain $\gamma_{Rb}$. We then fit the curves with the K MOT with eq. (S7) to obtain $a$ and consequently $\gamma_{K}(t)$. The values of $\gamma_K$ extracted in the two cases coincide within our experimental uncertainty.

\section{Chemical potential calculation}

To calculate the chemical potential of the thermal component $\mu$, we used the semi-classical distribution as described in~\cite{pethick_smith_2008}:

\begin{equation}
    {N_\text{th}}=g_3 (z(0)) \left( \frac{k_BT}{\hbar \omega} \right)^3 .
    \label{totaln}
\end{equation}
Where $T$ is the temperature, {$N_\text{th}$ is number of thermal atoms}, $\omega$ is the geometric mean of the trapping frequencies and $z$ is the fugacity defined as,

\begin{equation}
    z(\bf{r}) = e^{\left[ \mu - V(\bf{r}) \right]/k_BT} .
    \label{zr}
\end{equation}

Combining \ref{totaln} and \ref{zr} and considering that $V(0)=0$ we are able to obtain $\mu$ from our experimental data.

\section{The rate equation model}

A microscopic theory that is able to describe the out-of-equilibrium dynamics that governs the growth of the condensate is currently out of reach. In Refs. \cite{Jaksch_1997,Gardiner_1998,Gardiner2} a model is provided, however it is not always able to reproduce the experimental data and in some cases the parameters need to be adjusted up to one order of magnitude \cite{Jaksch_1997,Gardiner_1998,gaunt_superheated_2013,miesner_bosonic_1998}. The presence of the evaporation (therefore change in temperature and number of atoms) and of the dissipation, makes it even harder to describe our system with a microscopic theory.

Inspired by Refs \cite{Jaksch_1997,Gardiner_1998,Gardiner2,gaunt_superheated_2013} we have developed a two-mode rate equation model that can phenomenologically describe our data, as shown in Fig. 3 in the main text. Such model can provide useful information for the theory community interested in developing tools to describe the physics of out-of-equilibrium systems. 

{
As in the main text, we indicate with $N_0$ the number of atoms in the condensate and with $N_{th}$ the number of non-condensed atoms. As shown in Fig. \ref{model1}, the starting point of our model is the separation of the sample in the ground state energy level $\epsilon_0$, which contains the condensate, and the energy levels directly above it, which contain the thermal atoms, as done in \cite{gaunt_superheated_2013}. In our case, we simplify the problem reducing the energy band above the condensate to a single effective energy level $\epsilon_{th}$. Rigorously, $\epsilon_0$ should correspond to the time-dependent chemical potential of the BEC and also $\epsilon_{th}$ should dynamically change as the condensate grows or reduces. The understanding of the exact time dependence of the energy levels is probably the biggest challenge in developing an exact microscopic theory of these kind of systems, and goes beyond the scope of our work. Therefore we make the further approximation that $\epsilon_0$ and $\epsilon_{th}$ are time-independent. 

A rate equation that governs the growth of the population of a certain energy level in absence of dissipation has been developed in \cite{Jaksch_1997,Gardiner_1998,Gardiner2}:
\begin{equation}
    \do{N_n}=W(N)[(1-e^{(\epsilon_n-\mu_e)/k_BT})N_n+1],
    \label{eqmodel1}
\end{equation}
where $N_n$ is the population of the $n$-th level and $\epsilon_n$ its energy, which for the BEC coincides with the chemical potential of the condensate. $\mu_e$ is the \emph{equilibrium} chemical potential and $W(N)$ the growth coefficient. For the BEC state, the exponential implies that the steady state is reached when the chemical potential of the condensate equals the equilibrium chemical potential. In our model we have made the approximation that the energy levels are time independent. To account for the reaching of the steady state in absence of dissipation, we replace the exponential in eq. (\ref{eqmodel1}) with $t/t_f$, where $t_f$ is the time at which the steady state is reached and all the atoms are in the BEC.  

With the approximations made above, we can write write a two-mode rate equation model starting from the eq. (\ref{eqmodel1}) \cite{Jaksch_1997,Gardiner_1998,Gardiner2}:
\begin{eqnarray}
\dot{N}_0&=&\bar{W}\left[\left(1-\frac{t}{t_f}\right)N_0+1\right]-\bar{K}(\tilde{N}_{th}+1)-\gamma_K(t)N_0 \\ \nonumber
\dot{N}_{th}&=&-\bar{W}\left[\left(1-\frac{t}{t_f}\right)N_0+1\right]+\bar{K}(\tilde{N}_{th}+1)-[\gamma_K(t)+\gamma]N_{th},
\end{eqnarray}
{valid for $t\leq t_f$}. Here $\gamma_K$ is the dissipation rate coming from the K MOT and $\gamma$ the dissipation rate due to the evaporative cooling. The term proportional to $N_0$ in the square parenthesis and the one proportional to $\tilde{N}_{th}$ account for the bosonic stimulation \cite{Jaksch_1997,Gardiner_1998,Gardiner2}. The introduction of the effective number of thermal atoms $\tilde{N}_{th}$ accounts for the reduction of the non-condensed energy band to a single state. Indeed, if we were considering such state as populated by $N_{th}$ atoms, its  bosonic stimulation would be enhanced by a factor $N_{th}$. However, we need to take into account that the non-condensed energy band is not uniformly populated and that the more energetic levels are less involved in the exchange of particles. These two effects effectively reduce the bosonic stimulation. The amount of such reduction is difficult to evaluate, therefore $\tilde{N}_{th}$ is used as a free time-independent fit parameter.

\begin{figure}
    \centering
    \includegraphics[width=0.7\textwidth]{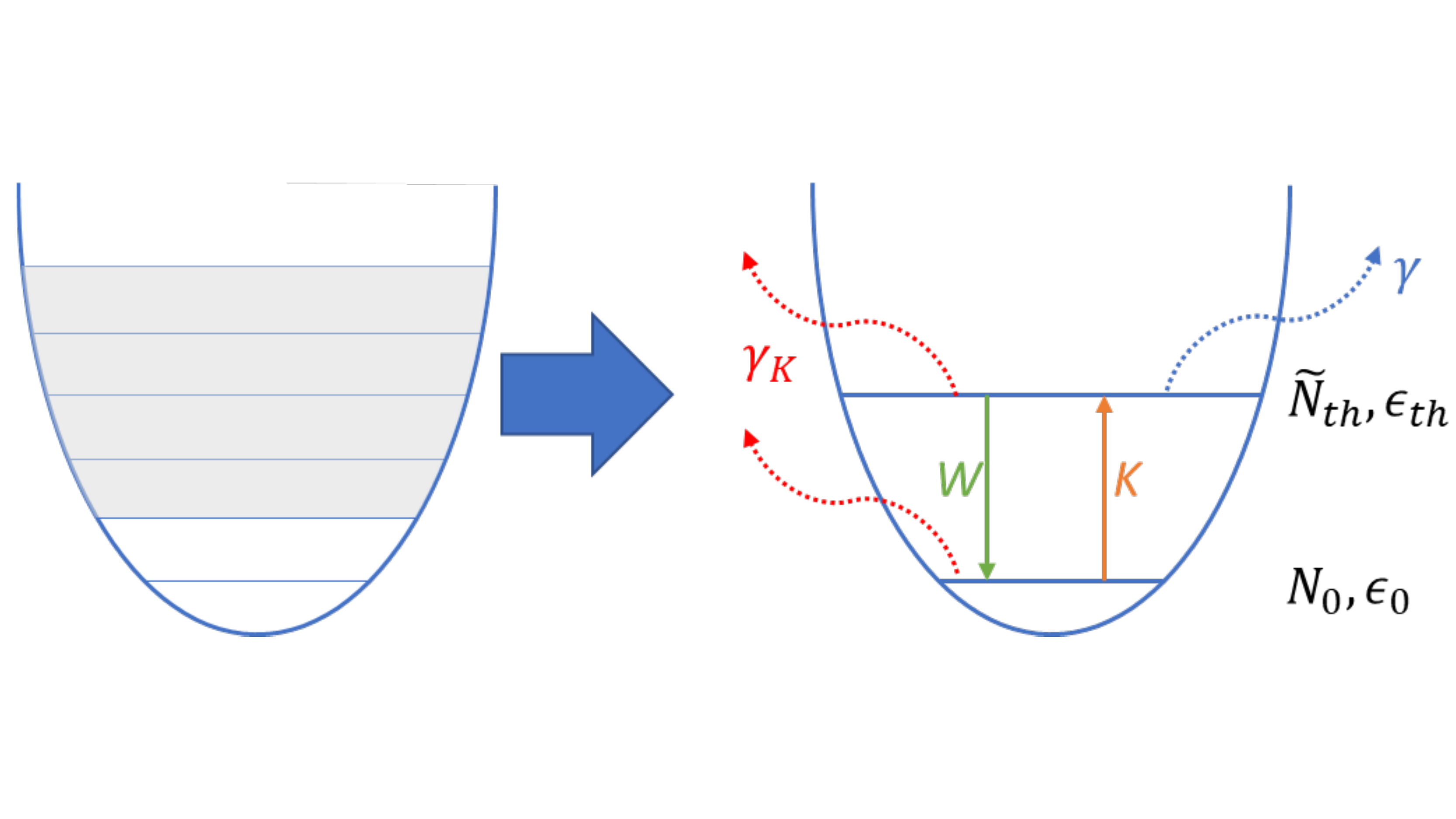}
    \caption{Pictorial representation of our rate equation model.}
    \label{model1}
\end{figure}

In general, the coefficients $W$ and $K$ are complicated functions of $T,\mu,N_0$ and $N$ \cite{Jaksch_1997,Gardiner_1998,Gardiner2} and an expression for them exists only for the case of constant temperature and infinite reservoir of atoms. Even in this case, as mentioned above, the coefficients can be wrong up to an order of magnitude. For this reasons, in our model we drop the functional dependence and treat them as constants having average values $\bar{W}$ and $\bar{K}$. In \cite{Jaksch_1997,Gardiner_1998}, the growth of the condensate is the result of the unbalancing between the two-body scattering process that transfers atoms from the thermal component to the BEC and the opposite process, without creation or absorption of quasiparticles. Using quantum kinetic theory, the authors found that 
\begin{equation}
\bar{W}\simeq e^{(\epsilon_{th}-\epsilon_0)/k_BT}\bar{K},    
\end{equation}
meaning that for a given temperature, the larger is the energy gap the harder is to excite an atom to the thermal component and the easier is to pump an atom into the condensate. The exploitation of this effect allows us to drive the system out of equilibrium. 

{In summary, in absence of dissipation our model is a simplification of the model in \cite{Jaksch_1997,Gardiner_1998,Gardiner2}. With respect to such model, we have maintained the following core features:
\begin{itemize}
\item The separation between condensed and thermal components
\item The bosonic stimulation, i.e., the fact that if $N$ bosons occupy a given state, the transition rates into that state are proportional to $(N + 1)$ 
\item The fact that the rates $W$ and $K$ are not equal
\item The fact that the growth of the condensate slows down and eventually stops as the target or equilibrium configuration is approached. In \cite{Jaksch_1997,Gardiner_1998,Gardiner2} this is done with the exponential term in S9, which decreases and finally stops the growth of the condensate once $\epsilon(t)=\mu_{e}$. In our model we have simplified this using the linear function $t/t_f$. We have verified that the use of a linear function instead of an exponential one does not substantially change the behaviour of the model.
\end{itemize}
The main modification we have made with respect to \cite{Jaksch_1997,Gardiner_1998,Gardiner2} are instead:
\begin{itemize} 
\item We have dropped the functional dependence of the coefficient $W$ and $K$. As explained, this is because, from previous comparisons of the model with experimental data, it was found that the coefficients calculated in \cite{Jaksch_1997,Gardiner_1998,Gardiner2} could be wrong up to one order of magnitude. For this reason, as a first approach, we have considered such coefficients as constant. Such approach is justified as in \cite{gaunt_superheated_2013} it has been proven that these coefficient do not change substantially to the first order if $\mu/T \ll k_B T$, as it is our case. In practice $W$ only depends on the elastic collision rate, which does not change substantially during our experiment. 
\item{We consider the non-condensed band as a single energy level with an effective population $\tilde{N}_{th}$ and we have dropped any time-dependence of the energy levels. As above, this simplification is dictated by the lack of a sufficiently precise microscopic model.}
\item To extend the model to describe our dissipative distillation, we have added the dissipative terms. 
\end{itemize} 
  }

\begin{figure}
    \centering
    \includegraphics[width=1\textwidth]{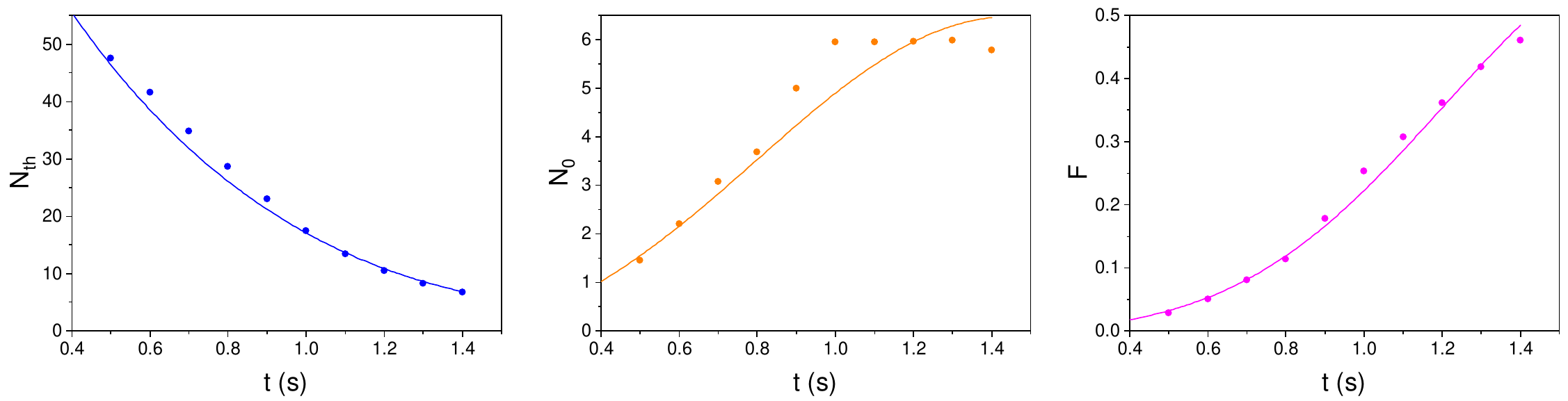}
    \caption{From left to right, evolution as a function of time of the number of atoms {(in thousands)} in the thermal component, the number of atoms in the BEC {(in thousands)} and the condensate fraction $F$. The points are the experimental data, the line are the results of our model.}
    \label{model2}
\end{figure}

Concerning the data shown in Fig.2, to derive $\bar{W}$ we fit the curve $N_0(t)$ in absence of dissipation with the solution of 
\begin{equation}
\dot{N}_0=\bar{W}\left[\left(1-\frac{t}{t_f}\right)N_0+1\right]    
\end{equation}
with $t_f=1.4$s the time at which the evaporation ends (please note the origin of time is shifted by 0.6s in this data analysis), obtaining $\bar{W}=3.45\pm0.1$ Hz. We derive $\bar{K}=0.255\pm0.03$ Hz from an exponential fit of the relaxation curve in Fig. 4 of the main text . The evaporation loss rate $\gamma=0.95\pm0.03$ Hz is obtained by fitting with an exponential the decay of the total number of atoms in absence of dissipation. Finally, $\gamma_K(t)$ is obtained as explained in the previous section. The only free parameter of our model is $\Xi=N_{th}/\tilde{N}_{th}$, which determines the effective occupation of the thermal energy band. We found that we can reproduce our data, as shown in Fig. \ref{model2}, by imposing $\Xi=25\pm5$. The result of our model is also reported as the dotted trajectory in the $\tau-F$ plane in Fig.3 of the main text. By increasing the dissipation rate $\gamma_K$ by a factor of four, leaving all the other parameters unchanged, we obtain the second trajectory that reproduces well the open circle data in Fig. 3. {Our data can be nicely reproduced using a constant value for $\Xi$, however our model could be extended to describe other kind of experiments, e.g. a quench in the number of atoms, implementing a time dependence on $\Xi$. }}

\bibliography{supplementary_bibl,main_bibl}